\documentclass[aps, pra, twocolumn, notitlepage, nofootinbib, longbibliography]{revtex4-1}

\usepackage{graphicx,amsmath,pstricks,float,subcaption,mathtools}

\begin{document}

\title{On the Likelihood of Observing Extragalactic Civilizations: Predictions from the Self-Indication Assumption}

\author{S. Jay Olson}
 \email{stephanolson@boisestate.edu}
 \affiliation{Department of Physics, Boise State University, Boise, Idaho 83725, USA}
 
\date{\today}

\begin{abstract}
Ambitious civilizations that expand for resources at an intergalactic scale could be observable from a cosmological distance, but how likely is one to be visible to us? The question comes down to estimating the appearance rate of such things in the cosmos --- a radically uncertain quantity. Despite this prior uncertainty, anthropic considerations give rise to Bayesian updates, and thus predictions. The Self-Sampling Assumption (SSA), a school of anthropic probability, has previously been used for this purpose. Here, we derive predictions from the alternative school, the Self-Indication Assumption (SIA), and point out its features. SIA favors a higher appearance rate of expansionistic life, but our existence at the present cosmic time means that such life cannot be too common (else our galaxy would long ago have been overrun). This combination squeezes our vast prior uncertainty into a few orders of magnitude. Details of the background cosmology fall out, and we are left with some stark conclusions. E.g. if the limits to technology allow a civilization to expand at speed $v$, the probability of at least one expanding cosmological civilization being visible on our past light cone is $1-\frac{v^3}{c^3}$. We also show how the SIA estimate can be updated from the results of a hypothetical full-sky survey that detects ``$n$'' expanding civilizations (for $n \geq 0$), and calculate the implied final extent of life in the universe. 
\end{abstract}

\maketitle

\section{Introduction}
One of the grand mysteries of the cosmos is the role of intelligent life. Is it a freak occurrence, rare and fleeting~\cite{klee2017}? Or is life destined to inherit and transform the universe on the largest scale~\cite{olson2014}?

To have an opinion on the matter is to have an opinion on the enabling factors --- the \emph{limits to practical technology}, and the \emph{appearance rate of ambitious life}. Severe bounds on either mean that life will forever be insignificant\footnote{``Insignificant'' in the sense that a physical description of the universe could be an excellent approximation without accounting for the activities of life.}. But the opposite case is quite possible.

It is not difficult to take the optimist position on technology. A minimal set of technologies required for this position are high-speed spacecraft that can travel between galaxies~\cite{fogg1988}, with the ability to self-replicate from available resources~\cite{freitas1980}. Current human technology is already developing $0.2 c$ interstellar spacecraft on a realistic budget~\cite{merali2016}, and initial studies suggest that the technology gap between interstellar and intergalactic space travel is slim~\cite{armstrong2013} --- the major difference is simply the travel time. Self-replicating technology is a greater challenge from our present position, but we are surrounded by numerous species of robust ``self-replicating aircraft'' that show the basic principle at work. 

The optimist position on technology implies that expanding cosmological civilizations~\cite{olson2014} are possible. Ambitious life, seeking to maximize access to resources, can embark on intergalactic expansion spanning billions of light-years, for nearly zero cost (due to use of self-replication). The more difficult question is that of the \emph{appearance rate}. If ambitious life is sufficiently rare, the universe on a large scale will hardly notice --- although such civilizations may become enormous, the cosmic acceleration will ensure they remain bounded and isolated in an otherwise sterile universe. A large appearance rate means that the universe will quickly become completely saturated with ambitious life.

Despite many decades of searches (from nearby stars~\cite{tarter2001} to as far as nearby galaxies~\cite{annis1999b,griffith2015}), no signatures of extraterrestrial intelligence have been demonstrated. Searches at deep cosmological distances may be forthcoming~\cite{lacki2019}, but until then, we are left with anthropic reasoning to estimate the appearance rate --- arguments that rely on the properties of our own existence~\cite{bostrom2002}. Two pieces of anthropic information are particularly important --- we have arrived at cosmic time $t_0 \approx 13.8$ Gyr, and we are apparently not within the domain of an ambitious civilization that is maximizing use of resources. 

There are two popular schools of anthropic probability~\cite{bostrom2002}. One is known as the Self-Sampling Assumption (SSA) --- it draws conclusions about the universe by reasoning as though our existence is a random draw from a ``reference class'' of similar life. It does not care how often civilizations like ours actually occur, but it favors conditions that make us more typical within the set of those that do occur. We have previously applied SSA reasoning to the problem of estimating the appearance rate (and observability) of expanding cosmological civilizations~\cite{olson2017a, olson2016, olson2018b}.

The other school is known as the Self-Indication Assumption (SIA). Rather than favoring our typicality within a reference class, SIA favors conditions that are more likely, in an absolute sense, to have produced us. SSA and SIA make different predictions, and the question of which is appropriate for cosmology has been debated for decades~\cite{dieks1992, olum2002, bostrom2003b}.

Our main goal here is to derive the basic predictions of SIA reasoning, for the appearance rate and observability of expanding cosmological civilizations. We will find the SIA approach has several attractive features. It is far less dependent on modeling of the relative appearance rate of life throughout cosmic time. It is far less dependent on one's prior assumptions about the magnitude of the appearance rate. The resulting predictions are simple, closed-form expressions. And they are somewhat more optimistic for detection than SSA-based estimates.

The SIA approach also lends itself well to the next step. Suppose that in the future, the full sky is surveyed to great distance, detecting ``$n$'' expanding cosmological civilizations. We show how this information updates our SIA estimate for the appearance rate. Based on this result, we can examine the original question above --- that of the ultimate role of life in the universe. At late cosmic times, how much of the universe will have been saturated by ambitious life?

We organize this paper in the following way: Section II reviews the basics of homogeneous cosmology with expanding civilizations. Section III discusses SSA and SIA in more detail, and how they are applied in the present context. Section IV discusses the issue of our prior uncertainty --- an essential starting-point for Bayesian reasoning. Section V derives our main result --- SIA estimates for the appearance rate and observability of expanding cosmological civilizations. Section VI shows how estimates are updated (with Bayes' theorem) after a search, and section VII addresses the question of the end state of life in the cosmos. Section VIII draws a direct comparison to previous SSA-based results, and section IX contains our concluding remarks.

\section{Expanding Cosmological Civilizations}

An expanding cosmological civilization is a hypothetical form of \emph{ambitious} life that has, as an instrumental goal, the control of as many resources as possible. It is difficult to imagine a more general final state of technological activity. In terms of utility, practically any final goal can be more fully realized with access to more resources~\cite{bostrom2014}. In terms of the energetics of life, an expanding cosmological civilization is a consequence of the maximum power principle~\cite{odum1995}. In terms of stability of large systems, the act of \emph{not} expanding requires all participants to agree not to expand, while initiating cosmic expansion requires only a single, isolated decision to launch one spacecraft that is capable of high speed and self-replication.   

Conservative assumptions on physics lead to a simple geometry for such a civilization --- expansion in all directions at some fraction of the speed of light, $v$.\footnote{We use units such that $c=1$, and velocities are expressed as a number less than 1.} The co-moving volume occupied by such a civilization at cosmic time $t$, having appeared at cosmic time $t'$ is that of a sphere, given by:
\begin{eqnarray}
v^3 \, V(t',t) = \frac{4 \pi}{3} \, v^3 \, \left( \int_{t'}^t \frac{1}{a(t'')} dt'' \right)^3
\end{eqnarray}
where $a(t)$ is the cosmic scale factor\footnote{We assume the following standard cosmology:  $\Omega_{\Lambda 0}=.692$, $\Omega_{r0}=9 \times 10^{-5}$, $\Omega_{m0}=1-\Omega_{r0} -\Omega_{\Lambda 0}$, $H_0 =.069 \, Gyr^{-1}$, giving $t_0 = 13.8$ Gyr.}. Note that $V(t',t_0)$ also represents the volume of space encompassed by our past light cone back at time $t'$.

At a large scale the universe is homogeneous, so the picture is that of spherical domains randomly appearing in space and expanding over a cosmic timescale, analogous to the geometry of a cosmological phase transition~\cite{guth1980,guth1981}. Indeed, it is more than an analogy, if life makes changes to the matter and radiation content. Intelligent life may literally be a way for the universe to quickly find and jump to a higher-entropy state, releasing heat in the process~\cite{olson2014}.

For our purposes, we regard the velocity of expansion, $v$, as the same constant for all expanding civilizations. The justification is that the difficulty of launching high-speed spacecraft explodes in a fairly narrow range of $v$. At $v=0.2c$, it is already practical for humanity~\cite{merali2016}. But $v=1$ is strictly impossible for any civilization, no matter how advanced, if our present understanding of physics is not completely misleading. Thus any ambitious civilization will likely encounter a very similar ``practical speed limit'' for their spacecraft, which is the main factor in their net expansion speed\footnote{The travel time between galaxies utterly dwarfs any other timescale in the process of expansion, e.g. the time required to reproduce between generations of spacecraft. Thus, if a civilization is ambitious, their net expansion speed will be very close to the speed of their intergalactic spacecraft. See appendix of~\cite{olson2018} for a detailed justification based on the spatial distribution of galaxies in the universe.}. Due to this approximately universal nature of $v$ (whatever its value may actually be), we refer to it as the \emph{dominant expansion velocity}. It is one of the two essential parameters in a cosmology with expanding civilizations.

The appearance rate of such civilizations (with units of appearances per $Gly^3$ of co-moving volume per $Gyr$ of cosmic time) is denoted by $f(t)$. Our prior assumptions about $f(t)$ are radically uncertain, but there is a simple way to isolate and express most of the uncertainty. We regard $f(t)$ as the product $f(t) = \alpha F(t)$, where $F(t)$ is a dimensionless time dependence (normalized to a maximum value of unity), roughly mirroring the cosmic rate of production of Earthlike planets~\cite{lineweaver2001}, with a lag of several Gyr to account for the long process of biological evolution~\cite{olson2017a} (see Figure 1). The parameter $\alpha$ (carrying the units of appearances per $Gly^3$ per $Gyr$) determines the overall scale, where most of the uncertainty lies. With a model for the background cosmology and $F(t)$ fixed, we regard a scenario of expanding cosmological civilizations as the pair of parameters $ \{ v, \alpha \} $. While $v$ is modestly uncertain, almost the entire subject of this paper boils down to finding an estimate for $\alpha$.

\begin{figure}
	\centering
	\includegraphics[width=0.49\textwidth]{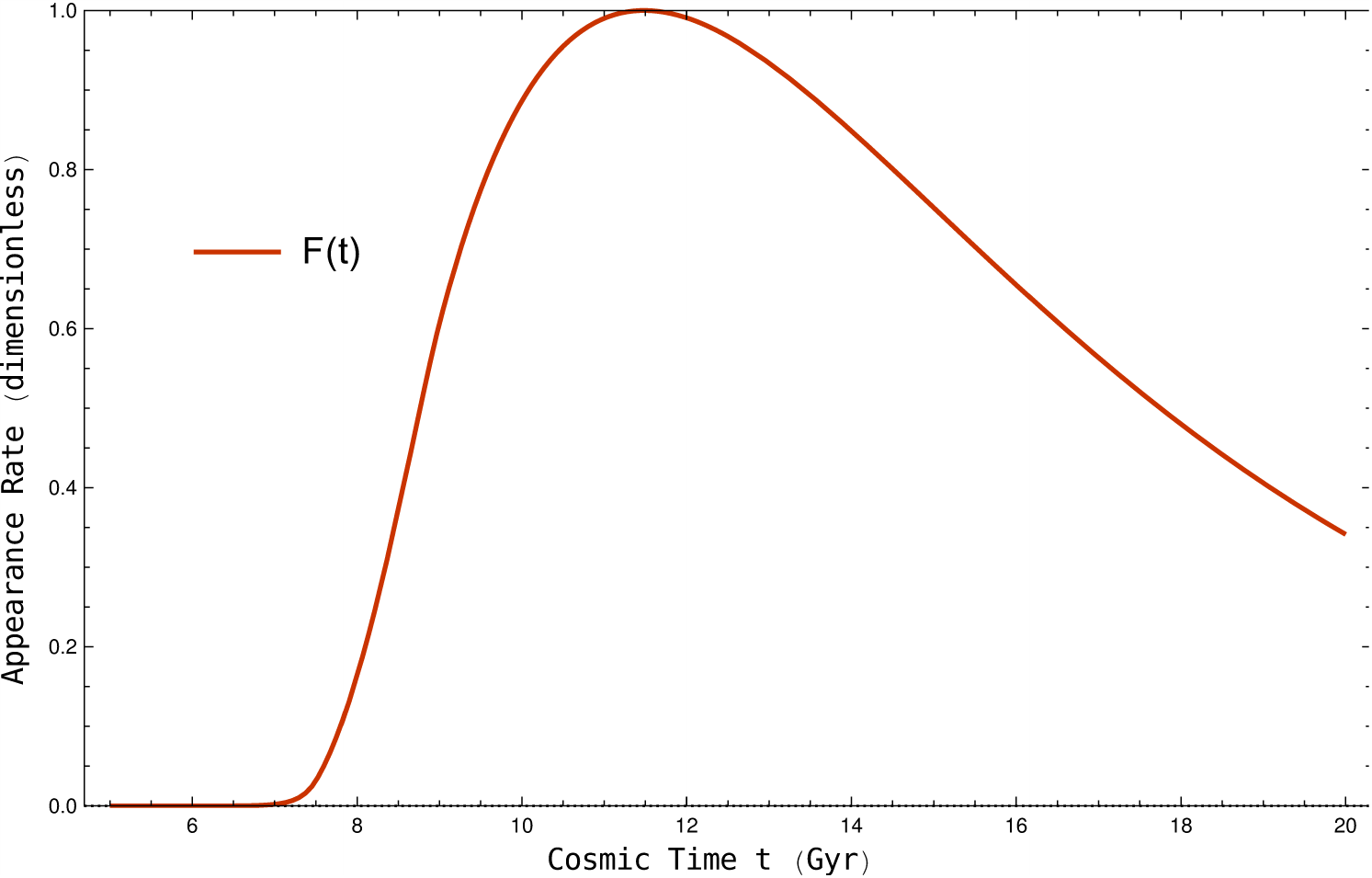}
	\caption{A model for the cosmic time-dependence of the appearance rate of advanced life, $F(t)$, normalized to a maximum value of unity.  The full appearance rate of ambitious expanding civilizations is given by $f(t)=\alpha F(t)$, where $\alpha$ has units of appearances per $Gly^3$ per $Gyr$. This model for $F(t)$ is effectively a model of the planet formation rate of the cosmos, with a lag of several Gyr to account for the evolution of intelligent life.}
\end{figure}

It is now convenient to define $s(t_0) = \int_{0}^{t_0} F(t) \, V(t,t_0) \, dt$. For the standard cosmology and the appearance rate function in Figure 1, $s(t_0) \approx 1268 \, Gly^3 \, Gyr$.  The fraction of the universe that remains \emph{un}saturated by expanding life at the present cosmic time is given by $g(t_0)$:
\begin{eqnarray}
g_{\alpha}(t_0) = e^{- \alpha v^3 s(t_0)} .
\end{eqnarray}
The (average) expected number of expanding cosmological civilizations that are present/visible on our past light cone is $E_{\alpha}(n)$:
\begin{eqnarray}
E_{\alpha}(n)=\alpha (1 -v^3) s(t_0),
\end{eqnarray}
and the probability that there is \emph{at least one} on our past light cone is then:
\begin{eqnarray}
p_{\alpha}(n \geq 1) = 1 - e^{-\alpha (1 -v^3) s(t_0)} .
\end{eqnarray}
The recurring factor of $(1 -v^3)$ expresses the fact that we count civilizations that have appeared within our past light cone, but \emph{not} within our ``past saturation cone'' (which expands into our past with speed $v$), since our position in space has apparently not been overtaken and saturated by an expanding civilization.

As a practical matter, if one can see only a fraction of the sky (due to the Zone of Avoidance or simply due to incomplete galaxy surveys), multiply $s(t_0)$ in the expressions for $E_{\alpha}(n)$ and $p_{\alpha}(n \geq 1)$ by the fraction of the sky that is visible.

\section{Self-Sampling and Self-Indication Assumptions}

The two most common schools of anthropic probability are known as the Self-Sampling Assumption (SSA) and the Self-Indication Assumption (SIA)~\cite{bostrom2002}. In this section, we give a quick review of the reasoning of each in the present context.

In SSA, we describe a ``reference class'' of civilizations that are analogous to humanity, and favor theories of the universe that make humanity more typical within the reference class. Essentially, we are viewing humanity as a random draw from a set of civilizations, and updating our understanding of the universe based on the properties of our draw.

There are multiple ways to arrive at the predictions of SIA --- here, we use a version called Full Non-Indexical Conditioning (FNC)~\cite{neal2006}, which agrees with SIA in the limit of observers who carry a significant amount of information\footnote{Predictions of FNC have been shown to differ from other presentations of SIA when ``observers'' are extremely simple beings, who carry only a few bits of information~\cite{armstrong2019}. We will not consider this possibility, and the results we obtain here can also be obtained using the older technique of modifying the prior to cancel the reference class. For this reason we use FNC and SIA interchangeably here.}. Effectively, SIA favors theories that make it more likely \emph{according to physical law} that we should exist here, in exactly our conditions. It works by saying the conditions most likely to have produced us, are also the conditions that produce many others like us. It makes no appeal to a reference class.

A good way to see the distinction is to consider the meaning of the probabilities appearing in a Bayesian application of SSA and SIA. The likelihood function, $P( \textrm{anthropic information} \, | \,  \textrm{theory n} )$ would, in SSA, represent the probability that a randomly selected member of the reference class sees \emph{anthropic information}, according to \emph{theory n}. In SIA, this quantity represents the probability that an observer who sees \emph{anthropic information} should appear at all, according to \emph{theory n}. The alternative (complement of \emph{anthropic information}) in SSA is that a different member of the reference class is selected. The alternative in SIA is that the observer who sees \emph{anthropic information} does not exist.

In the present context, the \emph{anthropic information} is humanity's cosmic time of arrival, $t_0$, while \emph{theory n} is the value of the appearance rate, $\alpha$.

The SSA approach we have previously used is to prescribe a reference class of human-stage civilizations that appear in galaxies that have not been saturated by ambitious life~\cite{olson2017a}. Then we can describe a cosmic time-of-arrival (TOA) distribution of such civilizations according to:
\begin{eqnarray}
p_{TOA}(t|\alpha) = \frac{F(t) g_{\alpha}(t)}{\int_0^\infty F(t') g_{\alpha}(t') dt'}.
\end{eqnarray}
Given SSA, one assumes that humanity is a random sample from this distribution. To get predictions, one specifies an assumption about humanity's ``typicality in time'' (whether were are at the average time of arrival, a one standard-deviation latecomer, etc.) and finds the corresponding value of $\alpha$ that matches $t_0$ with the assumption. This value of $\alpha$ is then used to get predictions according to equations 3 and 4.

The SIA approach we will develop below is more directly Bayesian. The (unnormalized) likelihood of our appearance here at cosmic time $t_0$ is:
\begin{eqnarray}
P(t_0 | \alpha, \gamma ) &=& \epsilon \, \gamma \, F(t_0) \, g_{\alpha}(t_0).
\end{eqnarray}
Here $\epsilon$ represents the probability of anthropic information that is irrelevant to our problem. That is, we carry a great deal of information in our memories that depends on local events that have happened on the Earth, and the probability of their occurrence is the fantastically small number $\epsilon$. It will cancel out with the normalization in Bayes' theorem. The new parameter $\gamma$ gives the appearance rate of \emph{human stage} civilizations (when multiplied by $F(t)$), and is related to $\alpha$ but carries is own uncertainty. It too is destined to fall out of the analysis, through its relationship to $\alpha$. Very significant to this approach is that the probability for humanity to occur at $t_0$ is proportional to $g_{\alpha}(t_0)$, the fraction of the universe that has not been overrun by ambitious life at the present cosmic time.

This Bayesian approach requires us to specify a prior pdf over $\alpha$.

\section{Prior Uncertainty}

Before any observations or anthropic considerations are made, we have essentially no ability to predict $\alpha$. Should $\alpha$ be of order $10$, or of order $10^{-10}$? To reflect this degree of uncertainty, we follow the reasoning of Tegmark~\cite{tegmark2014} and work exclusively with a prior pdf in $\alpha$, denoted $P(\alpha)$, that  assigns equal weight to each order of magnitude between $\alpha_{min}$ and $\alpha_{max}$. Explicitly, this is
\begin{eqnarray}
P(\alpha) = \left( \frac{1}{\log(\alpha_{max}/\alpha_{min})} \right) \frac{1}{\alpha}.
\end{eqnarray}

This kind of uncertainty tends to arise when a large number of modestly uncertain factors contribute to the value of $\alpha$. This can be seen to occur, for example, in an analysis of the Drake equation~\cite{vakoch2015} --- the product of seven rough estimates does not result in an ``order of magnitude estimate.'' It results in uncertainty that is inevitably spread over \emph{many} orders of magnitude~\cite{sandberg2018}.

Using the above prior, the question shifts to the endpoints. How many orders of magnitude should $P(\alpha)$ contain? Lacki has considered priors for the prevalence of life spanning $10^{122}$ orders of magnitude~\cite{lacki2016}, but in our context we only need the prior to stretch from about $\alpha_{min} = 10^{-6}$ to $\alpha_{max} = 10$ (appearances per $Glr^3$ per $Gyr$). Predictions from the SIA will not substantially change if the endpoints are extended further than that --- a surprising result to be justified below. In any case, the next section will derive closed-form results that are valid for any chosen values of $\alpha_{min}$ and $\alpha_{max}$, and then take the limit that the endpoints of the prior go to $\alpha_{min} \rightarrow 0$ and $\alpha_{max} \rightarrow \infty$.

\section{SIA Applied to Expanding Cosmological Civilizations: Predictions for Extragalactic SETI}

To update our prior, $P(\alpha)$, our cosmic time of arrival is particularly important. If the appearance rate $\alpha$ is too high, the universe would long ago have been completely saturated with expanding life. In other words, high values of $\alpha$ give $g_{\alpha}(t_0) \approx 0$, so there would be no opportunity for us to arise in an empty (by all appearances) galaxy.

SIA encodes our anthropic information by expressing the likelihood that we should appear at time $t_0$ in an untouched galaxy:
\begin{eqnarray}
P(t_0 | \alpha, \gamma ) &=& \epsilon \, \gamma \, F(t_0) \, g_{\alpha}(t_0).
\end{eqnarray}
As discussed in section III, $\epsilon$ represents the incredibly tiny probability of local events playing out for us in exactly the way they have, and $\gamma \, F(t_0)$ represents the cosmic appearance rate of \emph{human stage} life in unoccupied space, at the present cosmic time. However, $\gamma$ is not independent of $\alpha$, since all expanding civilizations presumably went through a stage of technological adolescence, similar to the current state of humanity. This implies $\gamma$ should be larger than $\alpha$ by some unknown factor. In other words, $\alpha = q \, \gamma$, where $q$ represents the fraction of human-stage civilizations that go on to become expanding cosmological civilizations. We expect that the question of whether a given human-stage civilization goes on to cosmological expansion depends almost entirely on local events, so $q$ should be independent of the cosmic time of arrival. So while $\alpha$ and $\gamma$ are not independent, we can regard $\alpha$ and $q$ as independent parameters, giving:
\begin{eqnarray}
P(t_0 | \alpha, q ) &=& \epsilon \, \frac{\alpha}{q} \, F(t_0) \, g_{\alpha}(t_0).
\end{eqnarray}
Since our prior assumptions about $\alpha$ and $q$ are independent of one another, and the above likelihood function factors, it is simple to marginalize over $q$, giving:
\begin{eqnarray}
P(t_0 | \alpha) &=& \epsilon \, \alpha \, F(t_0) \, g_{\alpha}(t_0).
\end{eqnarray}


Invoking Bayes' theorem, we then get the SIA prediction for $\alpha$, given that we have arrived at cosmic time $t_0$:
\begin{eqnarray}
P(\alpha | t_0 ) &=& \frac{\alpha \, g_{\alpha}(t_0) \, P(\alpha)}{\int_{\alpha_{min}}^{\alpha_{max}} \, \alpha' \, g_{\alpha'}(t_0) \, P(\alpha') \, d \alpha'} \\
&=&  \frac{s(t_0) \, v^3 \,  e^{(\alpha_{min}+ \alpha_{max} - \alpha)s(t_0)v^3 }}{e^{ \alpha_{max} s(t_0) v^3} -e^{  \alpha_{min} s(t_0) v^3} }.
\end{eqnarray}


So long as our prior is sufficiently spread out (and we avoid considering tiny values of the expansion speed $v$ --- more about this below), this result is well-approximated by taking the limit that $\alpha_{min} \rightarrow 0$ and $\alpha_{max} \rightarrow \infty$, giving:
\begin{eqnarray}
P(\alpha | t_0 ) &=& s(t_0) \, v^3 \, e^{-\alpha \, s(t_0) \, v^3}.
\end{eqnarray}

We can now use this posterior to get SIA predictions for the number of visible civilizations, and the probability at least one is visible, by using equations 3 and 4:

\begin{widetext}
\begin{eqnarray}
E(n) &=& \int_{\alpha_{min}}^{\alpha_{max}} \, E_{\alpha}(n) \, P(\alpha | t_0 ) \, d \alpha \\
 &=&   \left(1- v^3\right) \left(\alpha_{max} s(t_0) + \frac{1}{v^3} - \frac{s(t_0) (\alpha_{max}-\alpha_{min}) e^{\alpha_{max} s(t_0) v^3}}{e^{\alpha_{max} s(t_0) v^3}-e^{\alpha_{min} s(t_0) v^3}} \right) \\
 & \rightarrow & \frac{1}{v^3} - 1 
\end{eqnarray}
and
\begin{eqnarray}
p(n \geq 1) &=& \int_{\alpha_{min}}^{\alpha_{max}} \, p_{\alpha}(n \geq 1) \, P(\alpha | t_0 ) \, d \alpha \\
&=& 1-\frac{v^3 \left(e^{\alpha_{max} s(t_0)}-e^{\alpha_{min} s(t_0)}\right) e^{s(t_0) \left(v^3-1\right) (\alpha_{max}+\alpha_{min})}}{e^{\alpha_{max} s(t_0) v^3}-e^{\alpha_{min}s(t_0) v^3}}  \\
& \rightarrow & 1 - v^3 
\end{eqnarray}
\end{widetext}
where the arrows in the final lines indicate taking the limit of $\alpha_{min} \rightarrow 0$ and $\alpha_{max} \rightarrow \infty$. 

Since $E(n)$ diverges, one can see that the $\alpha_{max} \rightarrow \infty$ approximation is breaking down for tiny values of $v$. With smaller $v$, the posterior puts more weight on higher values of $\alpha$. At \emph{very} tiny values of $v$, it begins telling us that the appearance rate for expanding civilizations is greater than the formation rate of planets --- a non-physical manifestation of allowing $\alpha_{max} \rightarrow \infty$.

Reasonable values of $v$ avoid this. Already at $v = 0.1$, essentially all of the probability weight is well under $\alpha = 10$ appearances per $Gly^3$ per $Gyr$, and there is no problem with allowing $\alpha_{max} \rightarrow \infty$. In principle, $s(t_0)$ could also be made tiny to exacerbate the problem, but doing so would require a truly bizarre model of $F(t)$ to alter conclusions for $v \geq 0.1$. So long as estimates of $v$ and $s(t_0)$ are not completely shocking, the $\alpha_{max} \rightarrow \infty$ limit is an excellent approximation. This point is illustrated in Figure 2, where a direct comparison to the case of $\alpha_{min} = 10^{-5}$ and $\alpha_{max} = 10^{-1} $ is shown.

\begin{figure}
	\centering
	\includegraphics[width=0.49\textwidth]{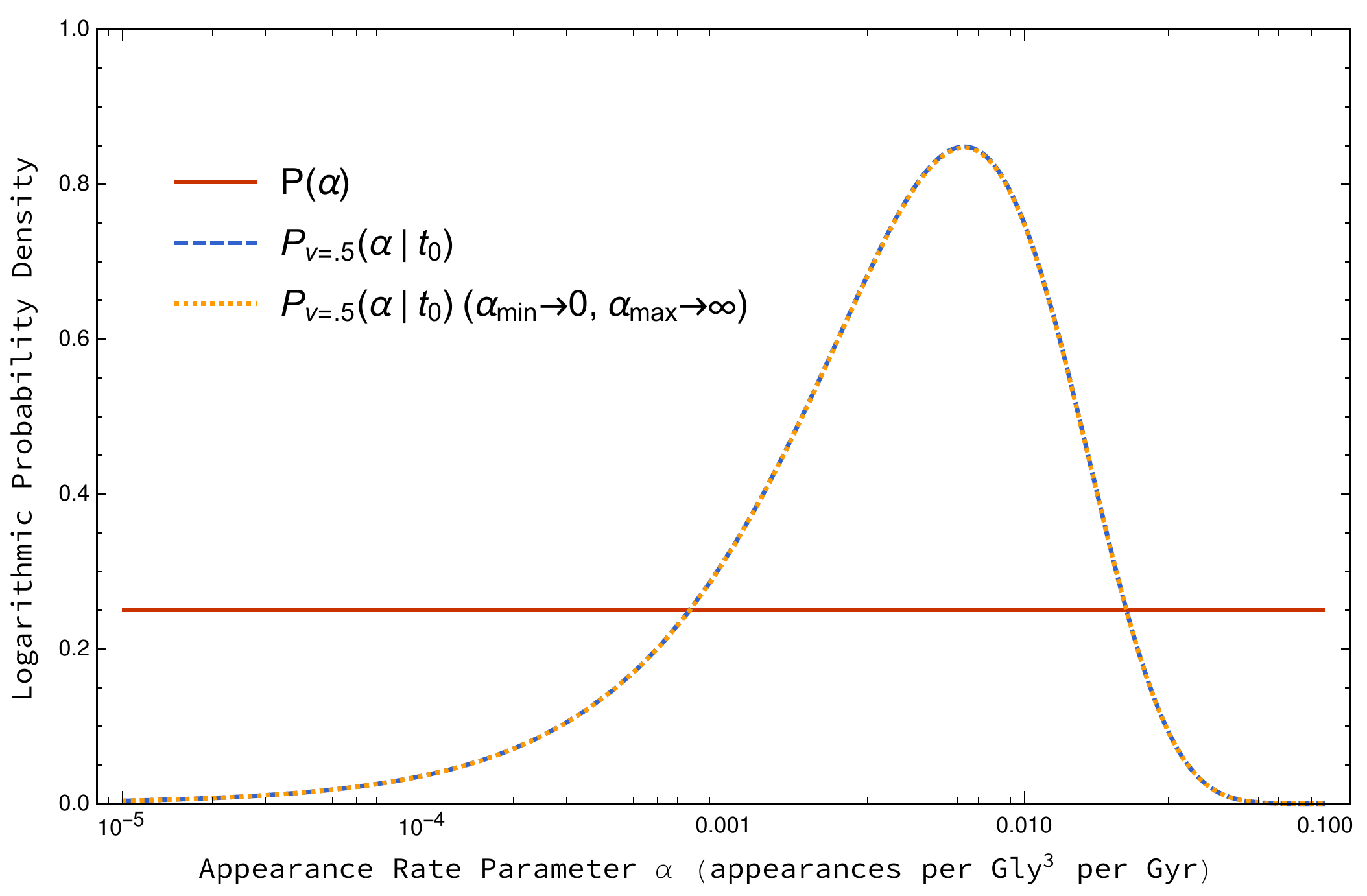}
	\caption{A prior pdf $P(\alpha)$ over the appearance rate $\alpha$ that stretches between $\alpha_{min} = 10^{-5}$ and $\alpha_{max} = 10^{-1}$, along with its anthropic SIA update $P(\alpha | t_0)$. Also depicted is the SIA update of a prior in the limit that $\alpha_{min} \rightarrow 0$ and $\alpha_{max} \rightarrow \infty$. There is no visible distinction --- SIA effectively does not care how spread out is the prior, beyond the first few orders of magnitude.}
\end{figure}

As noted in section II, it easy to account for the possibility that a survey covers only a fraction $FracSky$ of the sky. The SIA results for such a survey are:
\begin{eqnarray}
E(n) &=& FracSky \, \left( \frac{1}{v^3} - 1 \right) \\
p(n \geq 1) &=& \frac{1- v^3}{1 + \left( \frac{1}{FracSky} - 1  \right) v^3 }.
\end{eqnarray}
Results for values of $FracSky$ equal to $1$, $0.8$, and $0.35$ are shown in Figure 3. These values correspond respectively to the full sky, the full sky minus the Zone of Avoidance, and the coverage of the Sloan Digital Sky Survey (DR10)~\cite{Ahn2014}.

\begin{figure}
	\centering
	\includegraphics[width=0.49\textwidth]{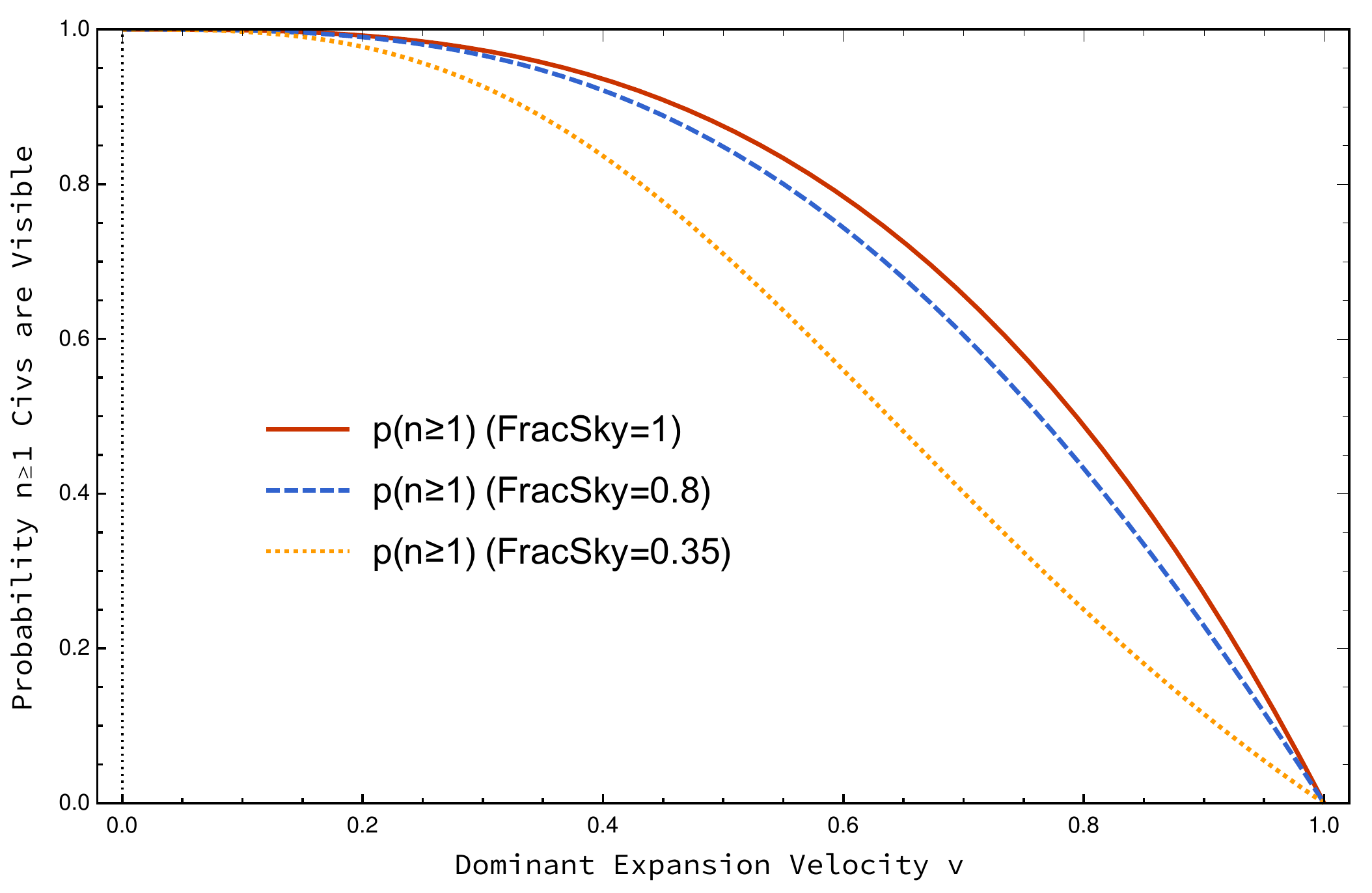}
	\caption{SIA estimated probability (as a function of the dominant expansion velocity $v$) that one or more expanding cosmological civilizations are visible on the full sky (FracSky = 1), the full sky minus the Zone of Avoidance (FracSky = 0.8), and the coverage of the Sloan Digital Sky Survey (FracSky = 0.35).  }
\end{figure}

\section{Updates Based on Search Results}

The Bayesian framework lends itself to taking the next step. Suppose that in the near future, a full-sky galaxy survey has been completed and discovers $n$ cosmological civilizations expanding at speed $v$. What would that do for our knowledge of $\alpha$? Such an update would be the final word on SIA estimates for $\alpha$, until a cosmological time has passed, allowing more information to slowly accumulate. 

We can perform such an update by noting the appearance of expanding civilizations is a Poisson process (appearances are independent events), and thus the probability to observe exactly $n$ is:
\begin{eqnarray}
p(n|\alpha) = \frac{E_{\alpha}(n)^n}{n!} e^{-E_{\alpha}(n)}.
\end{eqnarray}
Feeding this into Bayes' theorem to update our earlier SIA estimate gives:
\begin{eqnarray}
P(\alpha | n) = \frac{p(n|\alpha) P(\alpha | t_0 )}{\int p(n|\alpha') P(\alpha' | t_0 ) \, d \alpha' }.
\end{eqnarray}
Using the limit of $\alpha_{min} \rightarrow 0$ and $\alpha_{max} \rightarrow \infty$ for $P(\alpha | t_0 )$, this returns:
\begin{eqnarray}
P(\alpha | n) = \frac{s(t_0)}{ n!} \left(\alpha \, s(t_0) \right)^n \, e^{- \alpha \, s(t_0)}.
\end{eqnarray}
Note that this result is independent of $v$ --- the act of performing a definitive search cancels all velocity dependence from our knowledge of $\alpha$, even in the case of a null result, with an observation of $n=0$ civilizations.

The SIA estimate of $P(\alpha | t_0 )$ begins with uncertainty spread over about three orders of magnitude (for any proposed value of $v$). A null result ($n=0$) does not improve this uncertainty (except to eliminate dependence on $v$), but the greater the value of $n$, the less uncertainty we have. An observation of $n=1$ narrows the uncertainty to within two orders of magnitude. An observation of $n=10$ would narrow uncertainty to within a single order of magnitude. Figure 4 shows $P(\alpha | n)$ for several hypothetical values of $n$.

\begin{figure}
	\centering
	\includegraphics[width=0.49\textwidth]{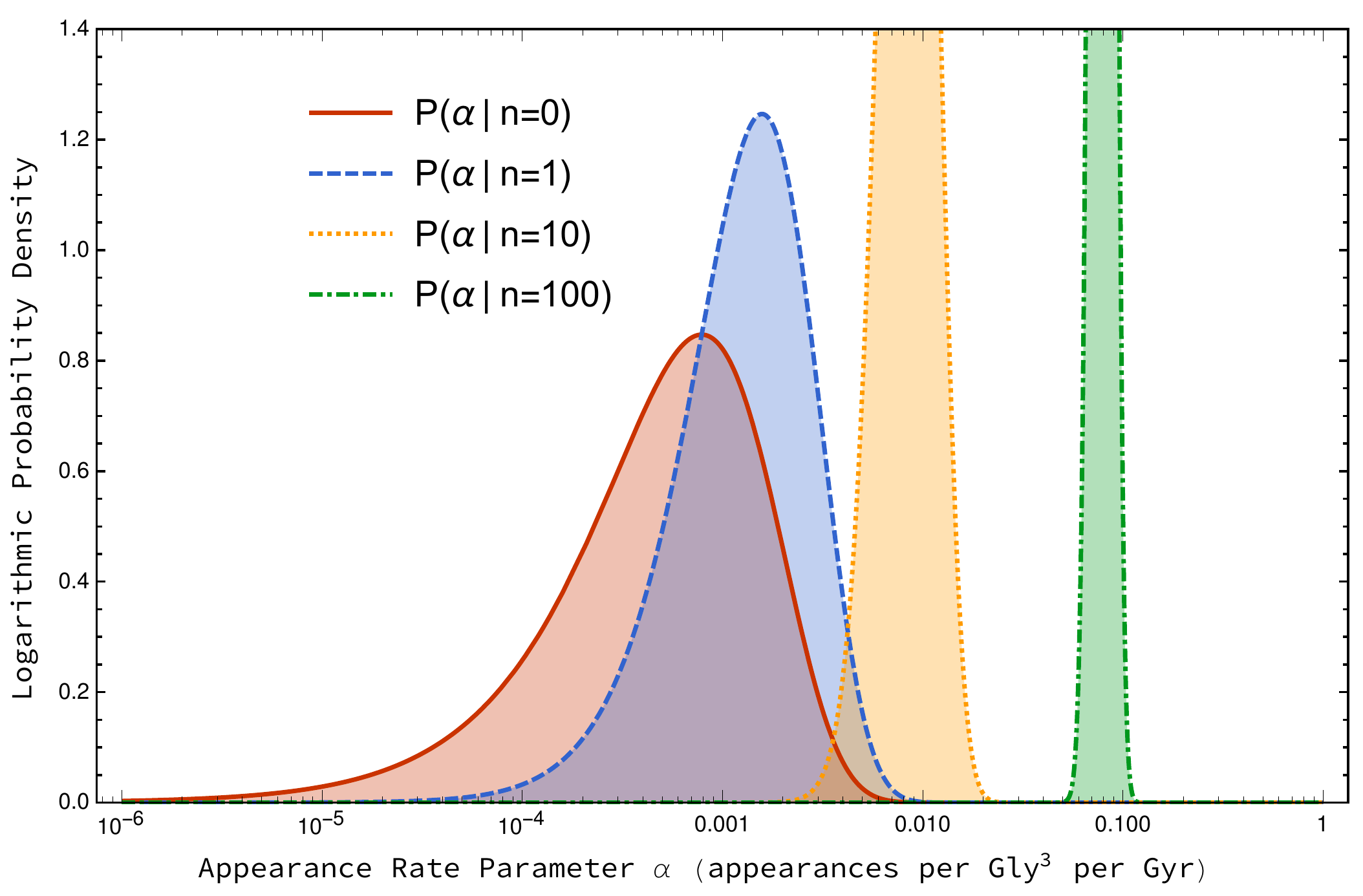}
	\caption{SIA estimate of $\alpha$ after update from a hypothetical full-sky survey that detects $n$ expanding civilizations, for $n=0$ (null result), $n=1$, $n=10$, and $n=100$.}
\end{figure}

\section{The Final Extent of life in the Cosmos}

After updating our SIA estimate to $P(\alpha | n)$ (from observing ``$n$'' expanding civs), we are in a position to address a fundamental question about the ultimate role of life in the universe.  

The fraction of the universe that remains forever untouched by life is $g_{\alpha}(t \rightarrow \infty)$, given by:
\begin{eqnarray}
 g_{\alpha}(t \rightarrow \infty) &=& e^{- \alpha v^3 s(\infty)}
 \end{eqnarray}
with $s(\infty) = \int_{0}^{\infty} F(t) \, V(t, \infty ) \, dt$ --- a finite quantity in the standard cosmology. With our model for $F(t)$, we get $s(\infty) \approx 207986 $.

We can then calculate its expected value, based on SIA and an observation of $n$ expanding civilizations at the current cosmic time, according to:
\begin{eqnarray}
E_{n}(g(t \rightarrow \infty)) &=& \int_{0}^{\infty} g_{\alpha}(t \rightarrow \infty) \, P(\alpha | n) \, d \alpha \\
&=& \left(  \frac{s(t_0)}{s(t_0) + s(\infty) \, v^3} \right)^{n + 1}.
\end{eqnarray}

\begin{figure}
	\centering
	\includegraphics[width=0.49\textwidth]{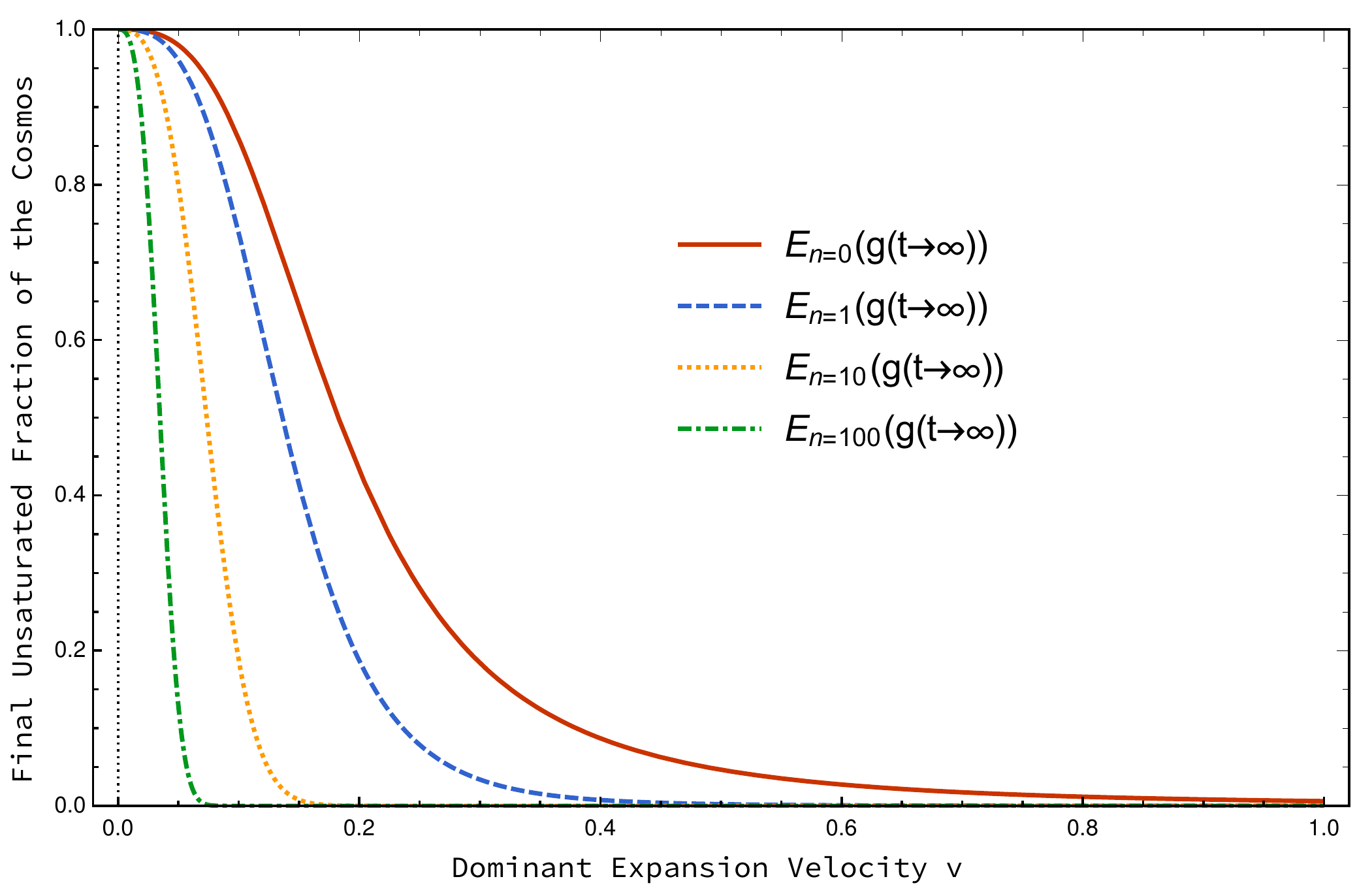}
	\caption{Expected fraction of the universe remaining untouched by expansionistic life at late cosmic time ($t \rightarrow \infty$), based on the updated SIA estimate of $\alpha$ after observing $n=0$, $n=1$, $n=10$, or $n=100$ expanding civilizations at the present cosmic time. If high-$v$ expansion is practical, the SIA expectation is simply that ambitious life will come to dominate the universe.}
\end{figure}

Figure 5 illustrates expected values of $g(t \rightarrow \infty)$. The conclusion is quite consistent for any value of ``$n$.'' If ambitious civilizations can expand at a middling to high fraction of the speed of light, then life will come to saturate most of the cosmos. This remains true even for the case of $n=0$ (a null search result) --- it is a nearly-unavoidable feature of using the SIA\footnote{Unlike the previous sections, this result depends on our model for $F(t)$ into the cosmic future (through $s(\infty)$). So although our conclusion is nearly independent of $n$, it is more vulnerable to major modeling changes.}.

\section{Comparison to SSA-based estimates}

The SIA estimate is different in character from the prior uses of SSA~\cite{olson2017a,olson2016,olson2018b}, offering major advantages and simplifications. It also introduces the well-known conceptual challenges connected to the SIA.

A drawback of the earlier SSA-based approach is that it requires a model of $F(t)$ throughout all time, including the extreme future, where such models are most uncertain. For example, the habitability of class M stars, with their extreme lifetimes, is controversial, and such considerations can heavily influence models for $F(t)$ in the far future~\cite{loeb2016}. It also contains a very counter-intuitive feature that life-hostile conditions at earlier times in the universe can \emph{increase} estimates for $\alpha$ and our probability to observe expanding civilizations\footnote{This can be understood as analogous to the Doomsday Argument. Assuming more life-hostile conditions at substantially earlier times in the universe means that we are closer to the earliest possible civilization, and this makes ``Doomsday'' (the appearance of expanding civilizations) more probable, to compensate.}~\cite{olson2017b}. The previous SSA approach has been criticized for a reliance on the idea of ``typicality in time.''~\cite{cirkovic2019} The results of the SSA approach also tend to be inconvenient to compute.

A simplified approach to using SSA is to set an anthropic bound --- a maximum plausible value of $\alpha$ that implies opportunities for human-stage life to appear are in rapid decline, due to displacement of galaxies by ambitious life~\cite{olson2018b}. This will imply bounds on $E(n)$ and $p(n \geq 1)$. While this approach is far easier to compute, it is merely a quick way to rule out high values of $\alpha$ --- it cannot make a true prediction without a prior over $\alpha$ that will control the result. 

The SIA approach developed in the previous sections answers nearly all of these drawbacks of the SSA. It does not rely on assumptions about humanity's typicality in time. It does not require a model of $F(t)$ that stretches into the distant future. It is easy to compute. Life-hostile conditions in the past do not paradoxically imply that life is easier to observe in the present. It makes very modest demands of our prior assumption about $\alpha$, requiring only that the prior is sufficiently spread out on a logarithmic scale to return closed-form, model-independent results. It is not vulnerable to the ``reference class problem'' of the SSA. And it makes true predictions, not merely setting bounds. 

It is also more optimistic than SSA-based predictions. If the dominant expansion speed is below about $v=0.8$, then SIA gives odds better than $50 \%$ that an expanding cosmological civilization is visible on our past light cone. To get similar predictions, earlier SSA estimates would require us to assume that humanity has appeared between one and two standard deviations later than the average time of arrival~\cite{olson2017a}.

On the other hand, the very fact that SIA predictions are so model-independent might actually be cause for skepticism. What is such model-independence saying? SIA is so strong that the background cosmology, the relative time dependence $F(t)$ for life, and the endpoints of the prior ($\alpha_{max}$ and $\alpha_{max}$)  are falling out of the key predictions because \emph{they are not important enough to matter.} This suggests that invoking SIA is a strong assumption indeed.

In fact, this is related to the ``Presumptuous Philosopher'' thought experiment, which was designed to showcase the unreasonable strength of SIA~\cite{bostrom2002,bostrom2003b}. In Presumptuous Philosopher, SIA so heavily favors conditions for plentiful life, that a philosopher is happy to go deeply into debt, placing bet after bet in favor of life, even in the face of strong experimental evidence piling up against him.  Invoking SIA, in the Presumptuous Philosopher scenario, is actually stronger than observation.

In the present case, the strength of SIA manifests in a slightly different way. SIA does not blindly favor ``more life'' --- it favors more life with exactly our anthropic information. Conditions that are too favorable for expansionistic life must be ruled out, because SIA favors conditions for us to appear \emph{at the present cosmic time, in an empty galaxy} --- too many expanding civilizations diminish the opportunity for that to occur. The strength of SIA is not manifesting in enormous estimates for life (the predicted values of $\alpha$ are, after all, rather tiny) --- it is manifesting in a very specific pdf, $P(\alpha | t_0 )$, no matter how conservatively spread out our prior may be. The amount of information gained by invoking SIA is enormous\footnote{The information gain (Kullback-Leibler divergence~\cite{kullback1959}) is infinite in the limit of $\alpha_{min} \rightarrow 0$ and/or $\alpha_{max} \rightarrow \infty$. Strictly speaking, these limits are unphysical, but they show us what the SIA is prepared to deliver on the basis of anthropic information alone.}. Our model could result in a presumptuous philosopher, if he is placing bets about $\alpha$ to occur within a few specific orders of magnitude.

\section{Conclusions}

To those familiar with SSA vs. SIA debates, it may not be surprising that an anthropic SIA model of extragalactic civilizations has two prominent features: Increased simplicity, and surprisingly high certainty. These are exactly the lessons of the SIA in two of the most famous anthropic thought experiments --- the Doomsday Argument and the Presumptuous Philosopher~\cite{bostrom2002}. 

Their presence here is stark. For dominant expansion velocity $v$ (whose value is a technological question), the ``bottom line'' is that the expected number of expanding cosmological civilizations that should be visible on our past light cone is $\frac{1}{v^3} - 1$, and the probability for us to see at least one is $1 - v^3$ --- this is true for any reasonable standard model cosmology, and any reasonable model for the relative appearance rate, $F(t)$. And no matter how conservatively spread out one's prior assumptions about the magnitude of the appearance rate --- take for example Lacki's $10^{122}$ orders of magnitude~\cite{lacki2016} --- the SIA posterior always narrows the uncertainty down to about three orders of magnitude. 

If one is convinced of the SIA, it difficult to avoid these conclusions. It could be the case that expanding civilizations have very little impact on their occupied galaxies, making them invisible from a cosmological distance. This would be difficult to square with the maximum power principle~\cite{odum1995}, and the fact that Dyson swarms are not a particularly exotic or difficult technology~\cite{dyson1960,lacki2019}. It could be that some expanding civilizations will deliberately adopt a strategy of ``stealth expansion'' until late cosmic times, to avoid influencing potential competitors~\cite{olson2018}. Or it could be that intergalactic expansion is physically impossible. None of these is a particularly compelling escape. In our opinion, the weakest link is the use of the SIA itself --- it would not be a huge surprise to learn that SSA or some other approach is more appropriate for cosmology.

\begin{acknowledgements}

I am grateful to Toby Ord, who asked me how the Self-Indication Assumption would change earlier conclusions.

\end{acknowledgements}

\bibliography{ref5}{}

\end{document}